\documentclass[preprint,showpacs,preprintnumbers,amsmath,amssymb]{revtex4-2}
\usepackage{bm}
\usepackage{amsmath}
\usepackage{amssymb}
\usepackage{amsfonts}
\usepackage{graphicx}
\begin{document}

\title{ \textbf{
    Three-nucleon force effects in 
    polarization transfers    
    from the doubly spin-polarized initial neutron-deuteron
    state to the outgoing neutron  
    in neutron-deuteron scattering} }

\author{H.~Wita{\l}a}
\affiliation{M. Smoluchowski Institute of Physics, 
Faculty of Physics, Astronomy and Applied Computer Science,
Jagiellonian University, PL-30348 Krak\'ow, Poland}

\author{J.~Golak}
\affiliation{M. Smoluchowski Institute of Physics, 
Faculty of Physics, Astronomy and Applied Computer Science,
Jagiellonian University, PL-30348 Krak\'ow, Poland}

\author{R.~Skibi\'nski}
\affiliation{M. Smoluchowski Institute of Physics, 
Faculty of Physics, Astronomy and Applied Computer Science,
Jagiellonian University, PL-30348 Krak\'ow, Poland}

\author{H.~Sakai}
\affiliation{RIKEN Nishina Center, 2-1, Hirosawa, Wako, 351-0198, Saitama,
  Japan }

\author{K.~Sekiguchi}
\affiliation{Department of Physics, Institute of Science Tokyo,
  Ookayama, Meguro, Tokyo 152-8551, Japan}

\date{\today}

\begin{abstract}
We discuss new spin observables presently
  accessible to  measurement in the proton-deuteron (pd) system, namely
  polarization   transfer coefficients from doubly
  spin-polarized initial state to the outgoing nucleon
  in the elastic nucleon-deuteron (Nd)        
  scattering and in the nucleon-induced deuteron breakup reactions.
  The sensitivity of these observables to three-nucleon force (3NF) effects is
  investigated and compared to sensitivities of the constituent
  standard single polarization transfer coefficients
  in the neutron-deuteron (nd) system.
  $K_{y,y}^{y'}$ in elastic
  nd scattering, for which large
  3NF effects, up to 40\%, have been found at higher energies, seems
  the most promising observable to measure.
\end{abstract}


\maketitle \setcounter{page}{1}

\section{Introduction}
\label{intro}

Reactions with polarized incoming particle(s) and/or polarizations of
outgoing particle(s) measured allow one to get more detailed information
about a transition matrix than unpolarized
processes~\cite{ohlsen1972,simonius1973}.
The simplest measurement of counting rates
with unpolarized initial state and no outgoing polarization measured
provides only
information on the magnitude of absolute values of
the transition matrix elements.
When one of the incoming particles is polarized, the measurement of the cross
section determines the so-called analyzing powers, which
give some   insights into relative phases between transition matrix
elements for
different spin projections of participating particles.
In this case complementary information about relative phases is obtained 
from a measurement  of a polarization of one of the outgoing
particles by providing the corresponding single spin transfer coefficient.
Polarizations
of different particles in the initial state lead to different
analyzing powers as well as to different single polarization transfer
coefficients. 

With the advancement of technology of ion sources and development of
sophisticated polarization techniques  more complex spin experiments
became available. Cross section measurements with both beam and target
polarized
provide spin correlation coefficients ~\cite{przevoski,kimsek},
which yield more insights into the transition matrix elements. Under
such initial conditions,
with both particles polarized in the initial state,  
 the measurement of a polarization
transfer to one of the outgoing particles offers the possibility to determine 
a new observable, completely different from single polarization
transfer coefficients,
when only one of the particles in initial state is polarized. The 
polarization transfer from such doubly spin-polarized
initial state will not only
provide more detailed information about transition matrix elements but
also will 
probably be sensitive in a  different way to the underlying dynamics.

With recent ubiquitous applications of chiral effective field theories to
nuclear structure
and reactions the question of importance of
three-nucleon forces in the nuclear Hamiltonian gained new impetus.
Rigorous three-nucleon (3N) Faddeev calculations allow one to obtain numerically
exact solutions of 3N Faddeev equations and predictions for observables
in 3N continuum reactions. 
In the present paper we investigate this new observable and its
sensitivity to the underlying dynamics in the neutron-deuteron
elastic scattering
and neutron-induced deuteron breakup reactions. In view of the planned
in the near future RIKEN measurement of the pd spin-correlation
coefficients ~\cite{kimsek} it seems
 relevant and timely to study properties of these new
observables and give some impetus for their measurement in the
future experiments.

The present limitations of experimental arrangements impose restrictions on
polarizations of the
incoming deuteron to the vector one in the $x$, $y$, and $z$ directions, 
and of the
incoming proton to the $y$ direction only. The polarization of the outgoing
proton  can be presently measured only 
in the $x$ and $y$ directions. (The right-handed Cartesian coordinate
system is defined according to the Madison convention~\cite{Madison1971}.)
Therefore we restrict our study to the following
double spin-polarization transfer coefficients $K_{d,n}^{n'}$:
$K_{x,y}^{x'}$, $K_{z,y}^{x'}$, and $K_{y,y}^{y'}$.
 We also investigate an
interesting case of  $K_{z,x}^{y'}$, for which both constituent single
polarization transfer coefficients vanish, as well as $K_{yy,y}^{y'}$ and
$K_{xz,y}^{y'}$ with the tensor polarized incoming deuteron. 

In Sec.~\ref{form} we briefly describe the 3N Faddeev formalism and give
 the definitions of the new observables. Predictions for different
polarization transfer
observables and a discussion of their sensitivity to the underlying dynamics
will follow in Sec.~\ref{results}. Finally, in Sec.~\ref{sumary} summary
and conclusions are given.

\section{Formalism}
\label{form}

Before defining  new observables,  we briefly outline, for
the reader's convenience,
 the main points of the 3N Faddeev formalism. 
 For details of the formalism  and of the numerical performance 
  we refer to ~\cite{glo96,wit88,hub97,book}.

Nucleon-deuteron (Nd) scattering with nucleons interacting
via nucleon-nucleon (NN) interactions $v_{NN}$ and a 3NF
$V_{123}=V^{(1)}+V^{(2)}+V^{(3)}$, is
described in terms of the breakup operator $T$ satisfying the
Faddeev-type integral equation~\cite{glo96,wit88,hub97}
\begin{eqnarray}
T\vert \phi \rangle  &=& t P \vert \phi \rangle +
(1+tG_0)V^{(1)}(1+P)\vert \phi \rangle + t P G_0 T \vert \phi \rangle \cr 
&+& 
(1+tG_0)V^{(1)}(1+P)G_0T \vert \phi \rangle \, .
\label{eq1a}
\end{eqnarray}
The two-nucleon (2N) $t$-matrix $t$ is the solution of the
Lippmann-Schwinger equation with the interaction
$v_{NN}$.   $V^{(1)}$ is the part of a 3NF which is 
symmetric under the interchange of
nucleons $2$ and $3$: $V_{123}=V^{(1)}(1+P)$, where
 the permutation operator $P=P_{12}P_{23} +
P_{13}P_{23}$ is given in terms of the transposition operators,
$P_{ij}$, which interchange nucleons $i$ and $j$.  The initial state 
$\vert \phi \rangle = \vert \vec {q}_0 \rangle \vert \phi_d \rangle$
describes the free motion of the nucleon  and the deuteron 
  with the relative momentum
  $\vec {q}_0$  and contains the internal deuteron wave function
  $\vert \phi_d \rangle$.
  $G_0$ is the free three-body resolvent.

 The amplitude for elastic scattering leading to the 
 final Nd state $\vert \phi ' \rangle$ is then given by~\cite{glo96,hub97}
\begin{eqnarray}
\langle \phi' \vert U \vert \phi \rangle &=& \langle \phi' 
\vert PG_0^{-1} \vert 
\phi \rangle  
 + \langle 
\phi'\vert  V^{(1)}(1+P)\vert \phi \rangle  \cr
&+& \langle \phi' \vert V^{(1)}(1+P)G_0T\vert  \phi \rangle +
\langle \phi' \vert PT \vert \phi \rangle ~,
\label{eq3}
\end{eqnarray}
while the  amplitude for the breakup reaction reads
\begin{eqnarray}
\langle  \vec p \vec q \vert U_0 \vert \phi \rangle &=&\langle 
 \vec p \vec q \vert  (1 + P)T\vert
 \phi \rangle ,
\label{eq3_br}
\end{eqnarray}
where the free  breakup channel state $\vert  \vec p \vec q \rangle $
is defined in terms of the  Jacobi (relative) momenta $\vec p$
and $\vec q$. 

We solve Eq.~(\ref{eq1a}) in the momentum-space partial-wave basis
$\vert p q \alpha \rangle$, determined by 
the magnitudes of the 
Jacobi momenta $p$ and $q$ and a set of discrete quantum numbers $\alpha$
comprising the 2N subsystem spin, isospin,
orbital and total angular momenta $s, t, l$ and $j$, as well as 
the spectator nucleon orbital
and total angular momenta with respect to the center of mass (c.m.) of the 2N
subsystem, $\lambda$ and $I$:
\begin{eqnarray}
\vert p q \alpha \rangle \equiv \vert p q (ls)j (\lambda \frac {1} {2})I (jI)J
  (t \frac {1} {2})T \rangle ~.
\label{eq4a}
\end{eqnarray}
The total 2N subsystem and spectator nucleon angular momenta
 $j$ and $I$, as well as isospins
$t$ and $\frac {1} {2}$, are finally 
coupled to the total angular momentum $J$ and isospin $T$ of the 3N system,
respectively.
In practice, a converged solution of Eq.~(\ref{eq1a})
using partial wave decomposition 
in momentum space at a given energy $E$ requires taking all 3N partial wave
states up to the 2N angular momentum $j_{max}=5$ 
and the 3N angular momentum $J_{max}=\frac{25}{2}$, with
the 3N force acting up to the 3N total
angular momentum $J=7/2$.
 Since in this study we omit completely
    the proton-proton (pp) Coulomb force
acting in the proton-deuteron system, 
which in elastic pd scattering has large effects only at
low energies and forward c.m. angles~\cite{delt2005,wita2024_1,wita2024_2},
the obtained results and conclusions are restricted at these energies
and angles to the nd system.

In the following we define new double spin-polarization transfer observables
for a reaction with two polarized particles in the initial state and
the polarization
of one of the outgoing particles measured: $\vec b(\vec a,\vec c)d$.
 In the case of elastic nd scattering the 
particles $b$ and $d$ will be the deuterons, and $a$ and $c$ the neutrons.
For details of the description of polarization states as well as reactions with
polarized particles we refer the reader to
\cite{ohlsen1972,simonius1973}. 

The density matrix of the initial state $\rho^{in}$ is given in terms of
polarization tensors of particles $a$, $t_{k_aq_a}$, and $b$, $t_{k_bq_b}$, by
~\cite{ohlsen1972,simonius1973}:
\begin{eqnarray}
  \rho^{in} &=& \frac {1} {(2s_a+1)(2s_b+1)}
  \sum_{k_bq_b} t_{k_bq_b}\tau_{k_bq_b}^{\dagger}
  \sum_{k_aq_a} t_{k_aq_a}\tau_{k_aq_a}^{\dagger} ~,
\label{eq1_a0}
\end{eqnarray}
where $\tau_{kq}$ are spherical tensor operators with $k=0, 1,\dots, 2s$, 
 $q=-k,-k+1, \dots, k$, and $s$ is particle's spin. 

The density matrix $\rho^{in}$ together with the transition operator $T$
for the reaction
$\vec b(\vec a,\vec c)d$, ($U$ of Eq.~(\ref{eq3}) for the nd elastic scattering
and $U_0$ of Eq.~(\ref{eq3_br}) for the deuteron breakup reaction),  
provides  the density matrix of the outgoing state 
 $\rho^{out}$
\begin{eqnarray}
\rho^{out} &=& T \rho^{in} T^{\dagger} ~.
\label{eq2_a0}
\end{eqnarray}

The polarization tensors of the outgoing particle $c$,
$t_{k_cq_c}(t_{k_aq_a}^a,t_{k_bq_b}^b)$, depend on the polarizations of
particles $a$ and $b$ in the initial state and are given by traces over
spin projections of the outgoing particles:
\begin{eqnarray}
  t_{k_cq_c}^c(t_{k_aq_a}^a,t_{k_bq_b}^b) &=&
  \frac {Tr(\rho^{out} \tau_{k_cq_c})} {Tr(\rho^{out})} \cr
  &=& \frac{\sigma^0} {\sigma} \sum_{k_aq_a,k_bq_b} t_{k_aq_a} t_{k_bq_b}
  \frac {Tr(T\tau_{k_bq_b}^{\dagger}\tau_{k_aq_a}^{\dagger}T^{\dagger}\tau_{k_cq_c})}
        {TT^{\dagger}} ~.
\label{eq3_a0}
\end{eqnarray}

Defining tensors of the polarization transfer:
\begin{eqnarray}
  t_{k_bq_b,k_aq_a}^{k_cq_c}(\vec b(\vec a, \vec c)d)  &\equiv&
\frac {Tr(T\tau_{k_bq_b}^{\dagger}\tau_{k_aq_a}^{\dagger}T^{\dagger}\tau_{k_cq_c})}
        {TT^{\dagger}} ~,    
\label{eq4_a0}
\end{eqnarray}
one has:
\begin{eqnarray}
  \sigma t_{k_cq_c}^c(t_{k_aq_a}^a,t_{k_bq_b}^b) &=&
  \sigma^0 \sum_{k_aq_a,k_bq_b} t_{k_aq_a} t_{k_bq_b}
 t_{k_bq_b,k_aq_a}^{k_cq_c}(\vec b(\vec a, \vec c)d) ~.
\label{eq5_a0}
\end{eqnarray}

If the terms with $k_a=0$ and/or $k_b=0$ in the sum in Eq.~(\ref{eq5_a0}) are 
separated, one gets:
\begin{eqnarray}
  \sigma t_{k_cq_c}^c(t_{k_aq_a}^a,t_{k_bq_b}^b) &=&
  \sigma^0 \lbrack
t_{00,00}^{k_cq_c}(b(a, \vec c)d) 
+  \sum_{k_b\ne 0 q_b}  t_{k_bq_b}
 t_{k_bq_b,00}^{k_cq_c}(\vec b(a, \vec c)d) \cr
&+&  \sum_{k_a\ne 0 q_a}  t_{k_aq_a}
 t_{00,k_aq_a}^{k_cq_c}(b(\vec a, \vec c)d) \cr
&+&  \sum_{\substack{k_a\ne 0 q_a\\k_b\ne 0 q_b}} t_{k_aq_a} t_{k_bq_b}
t_{k_bq_b,k_aq_a}^{k_cq_c}(\vec b(\vec a, \vec c)d)
\rbrack  ~.
\label{eq6_a0}
\end{eqnarray}

It should be emphasized that in (\ref{eq6_a0}) the cross section $\sigma$
depends on the polarization of the initial state and  differs for various 
 terms on the right hand side of this equation.
The first term is the induced polarization of particle $c$ in a reaction with
unpolarized initial state    
\begin{eqnarray}
  t_{k_cq_c}^{(0) c}(b(a, \vec c)d) &\equiv&
  t_{00,00}^{k_cq_c}(b(a, \vec c)d)
  ~.
\label{eq7_a0}
\end{eqnarray}
The second, third, and fourth  terms are contributions to the polarization of
the outgoing particle $c$ due to a single polarization transfer from a polarized
 particle $a$, due to a single polarization transfer from a polarized  
 particle $b$, and due to a double polarization transfer from doubly
 spin-polarized
 state of $a$ and $b$, respectively.
 The corresponding tensors of the polarization transfer,
 single: $t_{k_aq_a}^{k_cq_c}(b(\vec a, \vec c)d)$ or
 $t_{k_bq_b}^{k_cq_c}(\vec b(a, \vec c)d)$, and  double:
 $t_{k_bq_b,k_aq_a}^{k_cq_c}(\vec b(\vec a, \vec c)d)$,  are:
\begin{eqnarray}
 t_{k_aq_a}^{k_cq_c}(b(\vec a, \vec c)d)     &\equiv&
 t_{00,k_aq_a}^{k_cq_c}(\vec b(\vec a, \vec c)d)    \cr
 t_{k_bq_b}^{k_cq_c}(\vec b(a, \vec c)d)    &\equiv&
 t_{k_bq_b,00}^{k_cq_c}(\vec b(\vec a, \vec c)d)   \cr
 t_{k_bq_b,k_aq_a}^{k_cq_c}(\vec b(\vec a, \vec c)d)     &\equiv&
 t_{k_bq_b,k_aq_a}^{k_cq_c}(\vec b(\vec a, \vec c)d) 
  ~.
\label{eq8_a0}
\end{eqnarray} 
Thus the polarization of the outgoing particle $c$ is given by:
\begin{eqnarray}
  t_{k_cq_c}^c(t_{k_aq_a}^a,t_{k_bq_b}^b) &=&
  \frac {\sigma^0} {\sigma} \lbrack
  t_{k_cq_c}^{(0) c}(b(a, \vec c)d)
 + \sum_{k_a\ne 0 q_a}  t_{k_aq_a}
 t_{k_aq_a}^{k_cq_c}(b(\vec a, \vec c)d) \cr
&+&  \sum_{k_b\ne 0 q_b}  t_{k_bq_b}
 t_{k_bq_b}^{k_cq_c}(\vec b(a, \vec c)d) \cr
&+&  \sum_{\substack{k_a\ne 0 q_a\\k_b\ne 0 q_b}} t_{k_aq_a} t_{k_bq_b}
t_{k_bq_b,k_aq_a}^{k_cq_c}(\vec b(\vec a, \vec c)d)
\rbrack  ~.
\label{eq9_a0}
\end{eqnarray}

A direct calculation of polarization transfer tensors leads to the
double  spin-polarization transfer coefficients:
\begin{eqnarray}
  t_{k_bq_b,k_aq_a}^{k_cq_c}(\vec b(\vec a, \vec c)d)&=&
  \frac {1} { \sum_{m} |T_{m_cm_d}^{m_am_b}|^2 }
  \lbrack
  \sum_{\substack{m_c m_{c'}m_d\\m_a m_b m_{a'} m_{b'}}}
  T_{m_cm_d}^{m_am_b} (-1)^{q_b+q_a} \cr
 && \hat {s_b} (-1)^{s_b-m_{b'}}
  <s_b m_b s_b -m_{b'}|k_b-q_b > \cr
&&  \hat {s_a} (-1)^{s_a-m_{a'}}
  <s_a m_a s_a -m_{a'}|k_a-q_a >  T_{m_{c'}m_d}^{*m_{a'}m_{b'}} \cr
&& \hat {s_c} (-1)^{s_c-m_{c}}
  <s_c m_{c'} s_c -m_{c}|k_cq_c > 
  \rbrack  ~,
\label{eq10_a0}
\end{eqnarray}
and to the single polarization transfer coefficients: 
\begin{eqnarray}
  t_{k_aq_a}^{k_cq_c}(b(\vec a, \vec c)d)&=&
  \frac {1} { \sum_{m} |T_{m_cm_d}^{m_am_b}|^2 }
  \lbrack
  \sum_{\substack{m_c m_{c'}m_d\\m_a m_b m_{a'} }}
  T_{m_cm_d}^{m_am_b} (-1)^{q_a} \cr
&&  \hat {s_a} (-1)^{s_a-m_{a'}}
  <s_a m_a s_a -m_{a'}|k_a-q_a >  T_{m_{c'}m_d}^{*m_{a'}m_{b}} \cr
&& \hat {s_c} (-1)^{s_c-m_{c}}
  <s_c m_{c'} s_c -m_{c}|k_cq_c > 
  \rbrack  ~,
\label{eq11_a0}
\end{eqnarray}
where $s_i$ and $m_i$ are particles spins and their projections,  
$<s_1 m_1 s_2 m_2|k q >$ is a Clebsch-Gordan coefficient, and
$\hat s \equiv \sqrt{2s+1}$.

It should be emphasized that the polarization transfer tensors of
Eqs.~(\ref{eq10_a0}) and (\ref{eq11_a0})  are defined in the coordinate system
with  z-axis along the incoming particle's $a$ momentum.
Before comparing these observables to
data, they have to be transformed in indices $k_c$ and $q_c$ to
the coordinate system with the z'-axis along the outgoing particle's $c$ lab. 
momentum,  by performing a rotation around the y-axis through the outgoing
particle's $c$ 
lab. angle $\theta_c^{lab}$:
\begin{eqnarray}
  t_{k_bq_b,k_aq_a}^{k_cq_c}(\vec b(\vec a, \vec c)d) &=& \sum_{q'_c}
  D^{k_c}_{q'_c,q_c} (0 \theta_c^{lab} 0) 
  t_{k_bq_b,k_aq_a}^{k_cq'_c}(\vec b(\vec a, \vec c)d) ~,
\label{eq12_a0}
\end{eqnarray}
where $D^k_{q'q}(\alpha \beta \gamma)$ is the Wigner D-matrix for 
 a rotation specified by the Euler angles $(\alpha \beta \gamma)$
~\cite{brinksatch}. 

 In Appendix \ref{a1} we list all possible Cartesian double
 spin-polarization transfer
 coefficients to the outgoing neutron (altogether 81) and express
 them by the above spherical
polarization transfer tensors. Many of them (all underlined, altogether 41)
vanish due to parity conservation ~\cite{ohlsen1972}.

\section{Results and discussion}
\label{results}

To investigate the effects of a 3NF on polarization transfer coefficients,
particularly on polarization transfer coefficients from
double spin-polarized initial state,
we took the most precise chiral semilocal momentum space regularized (SMS)
potential of the Bochum group at N$^4$LO$^+$ order of chiral expansion
with the regulator
$\Lambda= 450$~MeV ~\cite{reinert2018}
and combined it with the N$^2$LO 3NF ~\cite{maris2021}. 
The contributions to that 3NF ~\cite{vankolck,epel2002} contain,
in addition to the 2$\pi$-exchange term,
two short-range contributions with the strength parameters c$_D$ 
and c$_E$ ~\cite{epel2020}. The latter two can be determined from the $^3$H 
binding energy and the Nd differential cross-section minimum at about
$E_{lab} = 70$~MeV i.e. the energy at which the effects of a 3NF are clearly
 noticeable in the Nd elastic-scattering cross
section \cite{wita1998,wita2001}. Namely, first the so-called
(c$_D$,c$_E$) correlation line is established, which for a particular chiral
NN potential combined with a N$^2$LO 3NF yields values of
(c$_D$,c$_E$) reproducing the $^3$H binding energy. Then, a fit to the
experimental data for the elastic Nd cross section is performed
and the values of both strengths, c$_D$ and c$_E$, are uniquely
determined. Taking pd elastic scattering cross sections at $E=65$~MeV of
Ref.~\cite{shim1982} this procedure determines the value of $
c_D = 2.10 \pm 0.24$ on the
correlation line. This justifies the values of $c_D=2.0$ and $c_E=0.2866$
from the correlation line, which we applied when solving Faddeev equations
for our combination of the NN potential  and this 3NF.

In Figs.~\ref{fig1}, ~\ref{fig2}, and ~\ref{fig3} we show predictions for
polarization transfer coefficients in elastic nd scattering
at three incoming neutron
lab. energies $E=10$, $70$, and $135$~MeV. First, at each energy we
present three polarization transfer coefficients
from doubly spin-polarized initial state, which
are potentially accessible to measurements in the pd system,
 namely: $K_{y,y}^{y'}$,
 $K_{x,y}^{x'}$, and $K_{z,y}^{x'}$. Three others, also potentially attainable for
 measurements: $K_{y,y}^{x'}$, $K_{x,y}^{y'}$, and $K_{z,y}^{y'}$, vanish
 for elastic scattering and in-plane breakup reaction. 
As was shown in ~\cite{ohlsen1972} conservation of parity requires, 
that in-plane  spin observables vanish unless the corresponding total number
of $x$ and $z$ indices is even. 
We also present predictions for $K_{z,x}^{y'}$, which
admittedly is not accessible experimentally at the moment, but is nevertheless
interesting, since despite the fact that it itself does not vanish,
 both single constituent transfer coefficients, from the 
 incoming neutron, $K_{x}^{y'}$, or deuteron, $K_{z}^{y'}$, to
 the outgoing neutron,  disappear.
  In order to provide information about the case when
     a doubly spin-polarized initial state contains
     the tensor polarized deuterons we
 show also the polarization transfer coefficients
 $K_{yy,y}^{y'}$ and $K_{xz,y}^{y'}$. 

 In each figure we show, in addition to the NN (black long-dashed line)  and
 NN+3NF (green long-dashed-dotted line) predictions for
 polarization transfer coefficients from doubly spin-polarized initial state
 $K_{d,n}^{n'}$ , 
 also results for the corresponding constituent
 single polarization transfer coefficients
 from the neutron to neutron $K_{0,p}^{p'}$
 (red solid line - NN and violet dotted line - NN+3NF) as well as  
  from the deuteron to neutron $K_{d,0}^{n'}$ (blue short-dashed line -  NN and
  orange short-dashed-dotted line - NN+3NF). We use this self-explaining
  notation for constituent single polarization transfer coefficients,
  $K_{0,n}^{n'}$ and $K_{d,0}^{n'}$, in all figures, placing them close to the
  corresponding lines. 

 Let us first discuss the polarization transfer coefficient $K_{y,y}^{y'}$
 from the double spin-polarized
 initial state to the neutron and its constituent single transfer
 coefficients, from the
 neutron or deuteron to the neutron, $K_{y}^{y'}$, 
  shown in Figs.~\ref{fig1}a-c.
 While at $E=10$~MeV $K_{y,y}^{y'}$ is significantly smaller than the single
 polarization transfer coefficients, at $E=70$ and $135$~MeV all three
 have comparable
 magnitudes. Effects of the 3NF, practically negligible at $E=10$~MeV, 
 become clearly visible at $E=70$,  with approximately the same magnitude
  (about $5 $\%)
 for all three polarization transfer coefficients. 
 With increasing energy the 3NF effects clearly grow, reaching
 at $E=135$~MeV about  $ 10 \%$ for single transfer coefficients
 but significantly
 more, about $ 40 \%$ for $K_{y,y}^{y'}$. 
 The 3NF effects  are  localized for the single polarization
 transfer coefficients in the region of the c.m. angles around 
 $\Theta_{c.m.}= 90^{o}$ while for  $K_{y,y}^{y'}$ this localization
 seems to be more strongly confined and shifted to c.m. angles around 
 $\Theta_{c.m.}= 110^o$. 

 For the constituent single polarization transfer coefficients from
 the neutron to neutron, $K_{y}^{x'}$, and
 from the deuteron to neutron,  $K_{x}^{x'}$, as well as for the $K_{x,y}^{x'}$,
 shown in Figs.~\ref{fig1}d-f, parity conservation requires
 vanishing of $K_{y}^{x'}$ ~\cite{ohlsen1972}. The nonvanishing single
 polarization transfer coefficient from the deuteron
  to the neutron, $K_{x}^{x'}$, reveals
 at $E=10$~MeV practically no 3NF effects in contrast to the  $K_{x,y}^{x'}$, for
 which effects of about $ 10 \%$ are seen  around
 $\Theta_{c.m.}= 120^o$. However, it must be emphasized that the magnitude
 of  $K_{x,y}^{x'}$ in this region of angles amounts only to
 about $ 0.1$.
    At $E=70$~MeV  $K_{x}^{x'}$ and $K_{x,y}^{x'}$ are practically
    not influenced by the 3NF with exception of a narrow
    region of c.m. angles around $140^o$, where both observables are small
    in magnitude.
  The 3NF effects become larger at
 $E=135$~MeV, changing even sign of $K_{x,y}^{x'}$, which is again small
 in that interesting region of angles. At this energy also $K_{x}^{x'}$ exhibits
 large 3NF effects, which increase the predictions based on 2N forces around
 $90^o$ and decrease them around $140^o$.

For $K_{z,y}^{x'}$ and corresponding constituent single polarization 
transfer coefficients 
from the neutron to neutron, $K_{y}^{x'}$, and
from the deuteron to neutron,  $K_{z}^{x'}$, shown in
Figs.~\ref{fig2}a-c, $K_{y}^{x'}$ again  vanishes.
At $E=10$~MeV only the polarization transfer coefficient from doubly
 spin-polarized initial state 
reveals effects of the 3NF of magnitude of about 
$10 \%$ in a region of backward angles. However, the
magnitude of $K_{z,y}^{x'}$  is only  of $\approx 0.1$ 
in this region of angles.  At $E=70$ and $135$~MeV this observable
reveals once more
effects of the 3NF at backward angles of the order of $\approx 10 \%$,
being however small in magnitude of $\approx 0.1$.
The single polarization transfer coefficient $K_{z}^{x'}$ shows
at these energies larger 
effects of the 3NF 
at $\Theta_{c.m.}\approx 90^o$, despite being small in magnitude at this
region of angles at $70$~MeV.

The polarization transfer coefficient from doubly spin-polarized
initial state $K_{z,x}^{y'}$, for which both constituent
 single polarization transfer coefficients,
from the neutron to neutron, $K_{x}^{y'}$, and
from the deuteron to neutron,  $K_{z}^{y'}$, vanish, represents
an interesting case (see Figs.~\ref{fig2}d-f). 
Even at $E=10$~MeV for $K_{z,x}^{y'}$ effects of the 3NF of the order of
$\approx 20 \%$ are seen at c.m. angles around $120^o$ and in the
backward region of angles.
These effects grow with increasing energy reaching at $E=135$~MeV
  about $50 \%$ raise with the magnitude of  $K_{z,x}^{y'}$
in this specific region of angles of about $0.15$.

    Changing in the doubly spin-polarized initial state
    the deuteron polarization
    to a tensor one leads again to practically no 3NF effects
    in double and single polarization transfers
    at $E=10$~MeV (Figs.~\ref{fig3}a and \ref{fig3}d). The 3NF effects
    start to appear at $E=70$~MeV (Figs.~\ref{fig3}b and \ref{fig3}e) and
    are clearly seen at $E=135$~MeV (Figs.~\ref{fig3}c and \ref{fig3}f)
    for both double $K_{yy,y}^{y'}$ and $K_{xz,y}^{y'}$
    as well as  all single polarization transfers $K_{yy,0}^{y'}$,
    $K_{0,y}^{y'}$, and $K_{xz,0}^{y'}$. Sizable effects for
    $K_{yy,y}^{y'}$ and $K_{xz,y}^{y'}$ occur for angles where these observables
    have large magnitudes of $\approx 0.4$.

We also studied the  polarization transfer coefficients
 from Figs.~\ref{fig1} and \ref{fig2} to the first outgoing neutron
 in exclusive in-plane deuteron
breakup reaction $d(n,n_1n_2)p$, choosing kinematically complete configurations
with largest cross sections. The first is a final-state-interaction (FSI)
geometry, in which two of the three outgoing nucleons, namely 1 and 3,
have equal momenta. Under this condition the relative energy of 
nucleons 1 and 3 is zero and they are strongly interacting in the two-body
partial wave state $^1S_0$. The second configuration is a quasi-free-scattering
 (QFS) geometry 
specified by a kinematical condition of vanishing laboratory energy of
the third nucleon. In both FSI(1-3) and QFS(1-2), the nucleons 1 and 2
are detected in coincidence at specific lab. angles $\theta_1^{lab}$ and
$\theta_2^{lab}$, what allows to unambiguously determine these
exclusive configurations
for a given lab. angle $\theta_1^{lab}$ of the outgoing nucleon 1.

In Figs.~\ref{fig4}-\ref{fig5} we show the predicted polarization
transfer coefficients at the three chosen
energies for FSI(1-3) geometry as a function of  $\theta_1^{lab}$.
At each energy in some region of $\theta_1^{lab}$ one has two solutions for
FSI(1-3), which correspond to different lab. angles
$\theta_2^{lab}$ of the outgoing nucleon 2, as well as to
 different lab. energies $E_1^{lab}=E_3^{lab}$.
 The second branch of solutions, which does not reach small lab. angles
 $\theta_1^{lab}$, corresponds to
solutions with small energies $E_1^{lab}=E_3^{lab}$.

At $E=10$~MeV not only all nonvanishing constituent
single polarization transfer coefficients but also
polarization transfer coefficients from doubly spin-polarized
 initial state are insensitive
 to action of the 3NF in all FSI(1-3) configurations
 (see Fig.~\ref{fig4} and \ref{fig5}).
At $E=70$~MeV only polarization transfer coefficients  from
 doubly spin-polarized initial state
 reveal effects of the 3NF in selected FSI(1-3) geometries.
 The polarization transfer coefficient
from the deuteron to the first neutron, $K_{z,0}^{x'}$ (see Fig.~\ref{fig5}b),
for which
changes by the 3NF are seen for FSI(1-3) geometries around
 $\theta_1^{lab}=20^o$, is an exception among single
polarization transfer coefficients.
The situation changes at higher energy $E=135$~MeV where both constituent single
polarization transfer coefficients as well as polarization
transfer coefficients from doubly 
spin-polarized initial state reveal large 3NF effects. Nonetheless, the
effects in polarization transfer coefficients from doubly 
spin-polarized initial state are the largest.

Finally, in Fig.~\ref{fig6} we show predictions for polarization transfer
coefficients at $E=135$~MeV in QFS(1-2) configurations.
Similarly to FSI(1-3), in some region of $\theta_1^{lab}$
one has two solutions for
QFS(1-2), which correspond to different lab. angles $\theta_2^{lab}$ 
 as well as to different lab. energies $E_1^{lab}$.
 The second branch of solutions, which does not reach small lab. angles
 $\theta_1^{lab}$, corresponds to
solutions with small energies $E_1^{lab}$.
It is interesting to see that in this geometry all polarization transfer
coefficients seem to be completely insensitive to the action of the 3NF.

    The experimental determination of different
    double spin-polarization transfer
    coefficients relies on a measurement of the outgoing particle's $c$
    polarization. According to (\ref{eq9_a0}), beside the induced polarization
    $t_{k_cq_c}^{(0)}$ given by  (\ref{eq7_a0}), three additional terms,
    whose magnitudes depend on the polarizations of particles in the
    initial state $t_{k_aq_a}$ and $t_{k_bq_b}$, as well as on the values of
    polarization transfer coefficients, contribute to this polarization.
    A successful measurement of a specific double spin-polarization transfer
    coefficient in a particular region of angles, requires sufficiently
    large values of the outgoing polarization as well as a significant
    contribution to it from the term containing this polarization
    transfer. All polarizations and spin observables are limited in magnitude
    by the conditions shown in ~\cite{ohlsen1972}. Since the contribution of the
    double polarization transfer to the outgoing polarization is governed
    by a product of the neutron and deuteron polarizations,  
    one expects its contribution to be smaller than from the single polarization
    transfers which come with one polarization only.
    Since the induced polarization comes without any polarization factor, one
    would expect it to be, when non zero, the dominant contribution to
    the outgoing polarization.

    One should emphasize that very probably the second requirement
            of the significant contribution to the final polarization is not
       so limiting for a measurement of the double spin-polarization transfer
            coefficients. Namely, such a  
            measurement is the last from a series of experiments,  
            in which only the double spin-polarization transfer is measured.
            All needed observables are determined already  in previously 
            performed measurements  of the neutron and deuteron analyzing
            powers, spin correlation coefficients, and  single
            neutron and deuteron polarization transfers.

For each double spin-polarization transfer 
    the question about feasibility of its measurement has to be answered
   separately by checking the above two requirements. In the following we do it
    for $K_{y,y}^{y'}$ at $E=135$~MeV by calculating the polarization of the
  outgoing neutron $P_{y'}^n$ and four contributions to it from
  the induced polarization $P_{y'}^{(0)}$, single neutron $K_{0,y}^{y'}$,
single deuteron $K_{y,0}^{y'}$, and double spin-polarization $K_{y,y}^{y'}$
    transfers:
\begin{eqnarray}
  P_{y'}^n &=& \frac {P_{y'}^{(0)} + p_y^n K_{0,y}^{y'}
      + \frac {3} {2} p_y^d K_{y,0}^{y'}
      + \frac {3} {2} p_y^n p_y^d K_{y,y}^{y'} } {1+p_y^nA_y^n
    + \frac {3} {2} p_y^d A_y^d
    + \frac {3} {2} p_y^n p_y^d C_{y,y}  } 
\label{eqqq_1}
\end{eqnarray}

To that end we have taken our  predictions 
based on the NN chiral potential only for the neutron $A_y^n$ and 
    deuteron $A_y^d$ analyzing powers, spin correlation coefficient $C_{y,y}$,
    induced neutron polarization $P_{y'}^{(0)}$, and single and double
    polarization transfers: $K_{0,y}^{y'}$, $K_{y,0}^{y'}$, $K_{y,y}^{y'}$.
    In Fig.~\ref{fig2}d-f we show the only nonvanishing component of
    the induced neutron polarization $P_{y'}^{(0)}$ by solid indigo (NN)
    and black dotted (NN+3NF) lines. To be realistic in our calculations 
    we took values 
    of the incoming neutron $p_y^n=p_y^p=0.2$ and deuteron $p_y^d=0.6$
    polarizations, achievable
 in present-day  polarized proton ~\cite{tateishi_ptarget1,sakaguchi_ptarget2}
          and deuteron    ~\cite{sekiguchi_dtarget} sources.

          In Fig.~\ref{fig7}a-d we show the final neutron polarization and
          contributions to it from each
          of the four terms in (\ref{eqqq_1}) for four sign combinations 
          of the incoming neutron and deuteron polarizations.
          The hierarchy of contributions obeys the expected pattern. 
          It starts with the largest induced polarization $P_{y'}^{(0)}$ part,
          followed by the
       single deuteron polarization transfer $K_{y,0}^{y'}$ contribution,
       followed by the single neutron polarization transfer $K_{0,y}^{y'}$
       term, and
       finally the double spin-polarization transfer part $K_{y,y}^{y'}$. 
       The sufficient values of the final neutron
       polarization ($\approx 0.6$) together with the reasonable 
          contribution of the term with a double spin-polarization transfer
          ($\approx 0.15)$  in the region
          of angles where large 3NF effects appear,  allow us to anticipate
          a successful measurement of $K_{y,y}^{y'}$.

\section{Summary and conclusions}
\label{sumary}

In this paper we propose  new spin observables which seem presently
  accessible to the present-day experimental facilities, namely
the polarization transfer coefficients from doubly spin-polarized initial state
of the pd elastic scattering to the outgoing proton.
Such a doubly spin-polarized
$\vec p \vec d$ state was used in a recent measurement of spin-correlation
coefficients at $E=135$ and $200$~MeV ~\cite{przevoski}
and will be also available in a soon-to-be conducted experiment
at RIKEN at $135$~MeV ~\cite{kimsek}.

We studied in the neutron-deuteron system  sensitivity to effects of the 3NF
 of three such polarization transfer coefficients
presently accessible to measurement:  $K_{y,y}^{y'}$,  $K_{x,y}^{x'}$, and
$K_{z,y}^{x'}$, and compared it with sensitivity of the constituent, single
 polarizations 
transfer coefficients from the neutron to neutron and from the deuteron
to neutron. A similar study was performed for the more difficult to
study experimentally polarization transfer coefficient $K_{z,x}^{y'}$,
exceptional
due to the fact that
both constituent single polarization transfer coefficients,
$K_{x}^{y'}$ and $K_{z}^{y'}$, vanish.
We also discussed $K_{yy,y}^{y'}$ and $K_{xz,y}^{y'}$ coefficients for
the reaction with tensor-polarized deuterons.
As a dynamical input in our study we took chiral N$^4$LO$^+$ SMS NN potential
of the Bochum group alone or combined with the chiral N$^2$LO 3NF.

We found that, similarly to the other elastic scattering observables,
effects of the 3NF are  significant at higher energies. 
The double spin-polarization transfer coefficient $K_{y,y}^{y'}$, for which
large 3NF effects were found at $E=135$~MeV is especially interesting. 
The 3NF effects for this
observable are  localized at c.m. angles $\approx 100^o$ and
are also by a factor of about two larger than for both constituent
single polarization
transfer coefficients $K_{y}^{y'}$. In that interesting region the magnitude
of $K_{y,y'}^{y'}$ is of the order of -0.3.
At this energy large 3NF effects were found for $K_{x,y}^{x'}$ and
$K_{z,y}^{x'}$. 
A similar pattern of 3NF effects has been found for polarization
transfer coefficients from the tensor polarized deuteron interacting
with polarized neutron.

Additionally, for the deuteron breakup reaction we investigated the
same double spin-polarization transfer coefficients in two chosen
kinematically complete configurations: FSI(1-3) and
QFS(1-2). While practically no 3NF effects are revealed in
quasi-free-scattering
configurations, significant 3NF effects are clearly seen for all four
double spin-polarization transfer coefficients in FSI(1-3) at $E=135$~MeV.
The most sizable effects are
localized in FSI(1-3) configurations around their production angle
$\theta_1^{lab} \approx 40^o$.

We are convinced that measurements of  these  double spin-polarization
transfer coefficients in the elastic pd
scattering and in some chosen QFS(1-2) and FSI(1-3) kinematically complete
 configurations of the
deuteron breakup will provide an interesting test for the 3NF effects
in these reactions.

\clearpage

\acknowledgements

This work was partly supported by the National Science Centre,
Poland under Grant
IMPRESS-U 2024/06/Y/ST2/00135, by JSPS KAKENHI Grant No. JP20H05636, Japan, and
 by JST ERATO Grant No. JPMJER2304, Japan.  
 It was also supported in part by the Excellence
Initiative – Research University Program at the Jagiellonian
University in Krak\'ow. 
The numerical calculations were partly performed on the supercomputers of
the JSC, J\"ulich, Germany.

\appendix

\section{Polarization transfer coefficients in elastic nd scattering
  from doubly spin-polarized initial state
  $\vec d \vec n$ to outgoing neutron $\vec n$:
  Cartesian observables expressed in terms of the
  spherical ones}
\label{a1}

For the nd elastic scattering with polarized neutron and deuteron
in the initial state  
$\vec d(\vec n,\vec{n'})d$ we denote the double   
spin-polarization transfer spherical tensors by $t_{k_dq_d,k_nq_n}^{k_{n'}q_{n'}}$
(defined in (\ref{eq4_a0})) 
 and the corresponding 
 double spin-polarization transfer Cartesian tensors
 by $K_{d,n}^{n'}$, defined analogously as in ~\cite{ohlsen1972} Eq.~(5.55):
 \begin{eqnarray}
   K_{j,i}^{l'}  &=& \frac{T r( T {\mathcal{P}}_j \sigma_i T^{\dagger}
     \sigma_{l'} ) }
   { Tr ( T T^{\dagger}) } \cr
   K_{jk,i}^{l'}  &=& \frac{T r( T {\mathcal{P}}_{jk} \sigma_i T^{\dagger}
     \sigma_{l'} ) }
   { Tr ( T T^{\dagger}) }   ~.
\label{eqqq_oh}
\end{eqnarray}   
 Using definitions
 of Ref.~\cite{ohlsen1972}, specifically relations between outgoing
 polarization, induced polarization and polarization transfers
  (Eqs.~(5.35) and (5.56)), 
 the following connections between double polarization
 Cartesian and spherical tensors  follow 
 (the underlined Cartesian tensors vanish for elastic scattering and
 in-plane breakup reaction due to parity conservation
 ~\cite{ohlsen1972}):

 1. $\underline{ K_{x,x}^{z'} } = \frac {2} {3} \times
 \frac {1} {2} \sqrt{\frac {3} {2} }
( t_{1-1,1-1}^{10} - t_{1-1,1+1}^{10}
             - t_{1+1,1-1}^{10} + t_{1+1,1+1}^{10}  )$

2. $K_{y,x}^{z'} = -\frac {2} {3} \times
             \frac {i} {2} \sqrt{ \frac{3} {2} }
             (t_{1-1,1-1}^{10} - t_{1-1,1+1}^{10}
             + t_{1+1,1-1}^{10} - t_{1+1,1+1}^{10}  )$

3. $\underline{ K_{z,x}^{z'} } = \frac {2} {3} \times
             \frac {\sqrt{ 3 }} {2}  (t_{10,1-1}^{10} - t_{10,1+1}^{10} )$

4. $K_{x,y}^{z'} =  - \frac {2} {3} \times
             \frac {i} {2} \sqrt{ \frac{3} {2} }
             ( t_{1-1,1-1}^{10} + t_{1-1,1+1}^{10} 
             - t_{1+1,1-1}^{10} - t_{1+1,1+1}^{10}  )$

5. $\underline{ K_{y,y}^{z'} } = -\frac {2} {3} \times
             \frac {1} {2} \sqrt{ \frac{3} {2} }
             (t_{1-1,1-1}^{10} + t_{1-1,1+1}^{10}
             + t_{1+1,1-1}^{10} + t_{1+1,1+1}^{10}  )$

6. $K_{z,y}^{z'} =  - \frac {2} {3} \times
             \frac {i \sqrt{3} } {2}  (t_{10,1-1}^{10} + t_{10,1+1}^{10} )$

7. $\underline{ K_{x,z}^{z'} } = \frac {2} {3} \times
             \frac {\sqrt{3}} {2}  (t_{1-1,10}^{10} - t_{1+1,10}^{10} )$

8. $K_{y,z}^{z'} = - \frac {2} {3} \times
             \frac {i \sqrt{3} } {2} (t_{1-1,10}^{10} + t_{1+1,10}^{10} )$

9. $\underline{ K_{z,z}^{z'} }= \frac {2} {3} \times
             \sqrt{ \frac{3} {2} }  t_{10,10}^{10} $


10. $K_{xx,x}^{z'} = \frac {1} {2} [
  -  ( t_{20,1-1}^{10} - t_{20,1+1}^{10} )
 +  \sqrt{ \frac {3} {2}}   ( t_{2-2,1-1}^{10} - t_{2-2,1+1}^{10}
                              + t_{2+2,1-1}^{10} - t_{2+2,1+1}^{10}  ) ]$
			      
11. $K_{yy,x}^{z'} = \frac {1} {2} [
  -  ( t_{20,1-1}^{10} - t_{20,1+1}^{10} )
 -  \sqrt{\frac {3} {2}}  ( t_{2-2,1-1}^{10} - t_{2-2,1+1}^{10}
                                + t_{2+2,1-1}^{10} - t_{2+2,1+1}^{10}  ) ]$
				
12. $\underline{ K_{xy,x}^{z'} }= - \frac {3} {2} \times
\frac {i} { \sqrt{6} }
                                ( t_{2-2,1-1}^{10} - t_{2-2,1+1}^{10}
                               - t_{2+2,1-1}^{10} + t_{2+2,1+1}^{10}  )$

13. $K_{xz,x}^{z'} = \frac {3} {2} \times 
                     \frac {1} { \sqrt{6} }  ( t_{2-1,1-1}^{10} - t_{2-1,1+1}^{10}
                            - t_{2+1,1-1}^{10} + t_{2+1,1+1}^{10}  )$

14. $\underline{ K_{yz,x}^{z'} } = - \frac {3} {2} \times \frac {i} { \sqrt{6} }
                            (  t_{2-1,1-1}^{10} - t_{2-1,1+1}^{10}
                                  + t_{2+1,1-1}^{10} - t_{2+1,1+1}^{10}  )$

15. $K_{zz,x}^{z'} =  ( t_{20,1-1}^{10} - t_{20,1+1}^{10} )$


16. $\underline{ K_{xx,y}^{z'} } =  \frac {1} {2} [
  {i}   ( t_{20,1-1}^{10} + t_{20,1+1}^{10} ) 
- {i} \sqrt{\frac {3} {2} } 
                                ( t_{2-2,1-1}^{10} + t_{2-2,1+1}^{10}
                              + t_{2+2,1-1}^{10} + t_{2+2,1+1}^{10}  ) ]$
			      
17. $\underline{ K_{yy,y}^{z'} } =  \frac {1} {2} [
   {i}   ( t_{20,1-1}^{10} + t_{20,1+1}^{10} ) 
 +  {i} \sqrt{\frac {3} {2} } 
                              (  t_{2-2,1-1}^{10} + t_{2-2,1+1}^{10}
                                + t_{2+2,1-1}^{10} + t_{2+2,1+1}^{10}  ) ]$
				
18. $K_{xy,y}^{z'} =  \frac {3} {2} \times \frac {1} { \sqrt{6} }
                                ( - t_{2-2,1-1}^{10} - t_{2-2,1+1}^{10}
                               + t_{2+2,1-1}^{10} + t_{2+2,1+1}^{10}  )$

19. $\underline{ K_{xz,y}^{z'} } = - \frac {3} {2} \times \frac {i} { \sqrt{6} }
                               (t_{2-1,1-1}^{10} + t_{2-1,1+1}^{10}
                            - t_{2+1,1-1}^{10} - t_{2+1,1+1}^{10}  )$

20. $K_{yz,y}^{z'} = - \frac {3} {2} \times \frac {1} { \sqrt{6} }
                            (t_{2-1,1-1}^{10} + t_{2-1,1+1}^{10}
                                  + t_{2+1,1-1}^{10} + t_{2+1,1+1}^{10}  )$

21. $\underline{ K_{zz,y}^{z'} } = -  {i} 
                                  ( t_{20,1-1}^{10} + t_{20,1+1}^{10} )$


22. $K_{xx,z}^{z'} = 
\frac {1} { 2} [- {\sqrt{2}}  t_{20,10}^{10}
  +  \sqrt{3}   ( t_{2-2,10}^{10} + t_{2+2,10}^{10}  ) ]$
			      
23. $K_{yy,z}^{z'} =  
\frac {1} { 2} [- {\sqrt{2}}  t_{20,10}^{10}
  - \sqrt{3}  ( t_{2-2,10}^{10} + t_{2+2,10}^{10}  ) ]$
				
24. $\underline{ K_{xy,z}^{z'} } = - \frac {3} {2} \times \frac {i} { \sqrt{3} }
                                ( t_{2-2,10}^{10} - t_{2+2,10}^{10}  )$

25. $K_{xz,z}^{z'} = \frac {3} {2} \times \frac {1} { \sqrt{3} }
( t_{2-1,10}^{10} - t_{2+1,10}^{10}  )$

26. $\underline{ K_{yz,z}^{z'} } = - \frac {3} {2} \times \frac {i} { \sqrt{3} }
                                    ( t_{2-1,10}^{10} + t_{2+1,10}^{10}  )$

27. $K_{zz,z}^{z'} =  {\sqrt{2}}  t_{20,10}^{10} $


28. $\underline{ K_{x,x}^{x'} }= \frac {2} {3} \times \frac {\sqrt{3}  } {4}
                            ( t_{1-1,1-1}^{1-1}
                             - t_{1-1,1-1}^{1+1}
                             - t_{1-1,1+1}^{1-1} + t_{1-1,1+1}^{1+1}
                             - t_{1+1,1-1}^{1-1} + t_{1+1,1-1}^{1+1}
                             + t_{1+1,1+1}^{1-1} - t_{1+1,1+1}^{1+1} )$

29. $K_{y,x}^{x'} = - \frac {2} {3} \times \frac {i\sqrt{3}} { 4 }
                             ( t_{1-1,1-1}^{1-1} - t_{1-1,1-1}^{1+1}
                                - t_{1-1,1+1}^{1-1} + t_{1-1,1+1}^{1+1}
                                + t_{1+1,1-1}^{1-1} - t_{1+1,1-1}^{1+1}
                                - t_{1+1,1+1}^{1-1} + t_{1+1,1+1}^{1+1} )$

30. $\underline{ K_{z,x}^{x'} }=  \frac {2} {3} \times
                                \frac {1} {2} \sqrt{\frac {3} {2} }
                                ( t_{10,1-1}^{1-1} - t_{10,1-1}^{1+1}
                   - t_{10,1+1}^{1-1} + t_{10,1+1}^{1+1} )$

31. $K_{x,y}^{x'} = - \frac {2} {3} \times \frac {i \sqrt{3}} { 4 }
                   ( t_{1-1,1-1}^{1-1} - t_{1-1,1-1}^{1+1}
                                  + t_{1-1,1+1}^{1-1} - t_{1-1,1+1}^{1+1}
                                  - t_{1+1,1-1}^{1-1} + t_{1+1,1-1}^{1+1}
                                  - t_{1+1,1+1}^{1-1} + t_{1+1,1+1}^{1+1} )$

32. $\underline{ K_{y,y}^{x'} }= \frac {2} {3} \times \frac {\sqrt{3}} { 4  }
                                  ( - t_{1-1,1-1}^{1-1} + t_{1-1,1-1}^{1+1}
                                 - t_{1-1,1+1}^{1-1} + t_{1-1,1+1}^{1+1}
                                 - t_{1+1,1-1}^{1-1} + t_{1+1,1-1}^{1+1}
                                 - t_{1+1,1+1}^{1-1} + t_{1+1,1+1}^{1+1} )$

33. $K_{z,y}^{x'} = - \frac {2} {3} \times \frac {i} {2} \sqrt{\frac {3} {2} }
                                 ( t_{10,1-1}^{1-1} - t_{10,1-1}^{1+1}
                      + t_{10,1+1}^{1-1} - t_{10,1+1}^{1+1} )$

34. $\underline{ K_{x,z}^{x'} }= \frac {2} {3} \times
                      \frac {1} {2} \sqrt{\frac {3} {2} }
                      ( t_{1-1,10}^{1-1} - t_{1-1,10}^{1+1}
                    - t_{1+1,10}^{1-1} + t_{1+1,10}^{1+1} )$

35. $K_{y,z}^{x'} = - \frac {2} {3} \times \frac {i} {2} \sqrt{\frac {3} {2} }
                    ( t_{1-1,10}^{1-1} - t_{1-1,10}^{1+1}
                      + t_{1+1,10}^{1-1} - t_{1+1,10}^{1+1} )$

36. $\underline{ K_{z,z}^{x'} }= \frac {2} {3} \times \frac {\sqrt{3} } { 2 }
                      ( t_{10,10}^{1-1} - t_{10,10}^{1+1} )$


37. $K_{xx,x}^{x'} =   \frac {1} { 2 } [
  - \frac {1} { \sqrt{2} }  ( t_{20,1-1}^{1-1} - t_{20,1-1}^{1+1}
                                   - t_{20,1+1}^{1-1} + t_{20,1+1}^{1+1} )
 +  \frac {\sqrt{3}} {2} ( t_{2-2,1-1}^{1-1} - t_{2-2,1-1}^{1+1}
                              - t_{2-2,1+1}^{1-1} + t_{2-2,1+1}^{1+1} 
	                      + t_{2+2,1-1}^{1-1} - t_{2+2,1-1}^{1+1}
                              - t_{2+2,1+1}^{1-1} + t_{2+2,1+1}^{1+1} ) ] $

38. $K_{yy,x}^{x'} =  \frac  {1} { 2 } [
  - \frac {1} { \sqrt{2} }  ( t_{20,1-1}^{1-1} - t_{20,1-1}^{1+1}
                                   - t_{20,1+1}^{1-1} + t_{20,1+1}^{1+1} )
 - \frac {\sqrt{3}} {2} ( - t_{2-2,1-1}^{1-1} + t_{2-2,1-1}^{1+1}
                                + t_{2-2,1+1}^{1-1} - t_{2-2,1+1}^{1+1} 
	                        - t_{2+2,1-1}^{1-1} + t_{2+2,1-1}^{1+1}
                                + t_{2+2,1+1}^{1-1} - t_{2+2,1+1}^{1+1} )$

39. $\underline{ K_{xy,x}^{x'} }= - \frac {3} {2} \times \frac {i} { 2 \sqrt{3} }
                                ( t_{2-2,1-1}^{1-1} - t_{2-2,1-1}^{1+1}
                                   - t_{2-2,1+1}^{1-1} + t_{2-2,1+1}^{1+1} 
	                           - t_{2+2,1-1}^{1-1} + t_{2+2,1-1}^{1+1}
                                   + t_{2+2,1+1}^{1-1} - t_{2+2,1+1}^{1+1} )$

40. $K_{xz,x}^{x'} = \frac {3} {2} \times \frac {1} { 2 \sqrt{3} }
                                   ( t_{2-1,1-1}^{1-1} - t_{2-1,1-1}^{1+1}
                              - t_{2-1,1+1}^{1-1} + t_{2-1,1+1}^{1+1} 
	                      - t_{2+1,1-1}^{1-1} + t_{2+1,1-1}^{1+1}
                              + t_{2+1,1+1}^{1-1} - t_{2+1,1+1}^{1+1} )$

41. $\underline{ K_{yz,x}^{x'} }= - \frac {3} {2} \times \frac {i} { 2 \sqrt{3} }
                              ( t_{2-1,1-1}^{1-1} - t_{2-1,1-1}^{1+1}
                                   - t_{2-1,1+1}^{1-1} + t_{2-1,1+1}^{1+1} 
	                           + t_{2+1,1-1}^{1-1} - t_{2+1,1-1}^{1+1}
                                   - t_{2+1,1+1}^{1-1} + t_{2+1,1+1}^{1+1} )$

42. $K_{zz,x}^{x'} =  
                   \frac {1} {  \sqrt{2} }  ( t_{20,1-1}^{1-1} - t_{20,1-1}^{1+1}
                                   - t_{20,1+1}^{1-1} + t_{20,1+1}^{1+1} )$ 


43. $\underline{ K_{xx,y}^{x'} } = \frac {1} {2} [
 \frac {i} { \sqrt{2} }
                                   ( t_{20,1-1}^{1-1} - t_{20,1-1}^{1+1}
 + t_{20,1+1}^{1-1} - t_{20,1+1}^{1+1} )
                                    - \frac {i \sqrt{3}} { 2 }
                             ( t_{2-2,1-1}^{1-1} - t_{2-2,1-1}^{1+1}
                              + t_{2-2,1+1}^{1-1} - t_{2-2,1+1}^{1+1} 
	                      + t_{2+2,1-1}^{1-1} - t_{2+2,1-1}^{1+1}
                              + t_{2+2,1+1}^{1-1} - t_{2+2,1+1}^{1+1} ) ]$

44. $\underline{ K_{yy,y}^{x'} } =  \frac {1} {2} [
 \frac {i} { \sqrt{2} }
                                   ( t_{20,1-1}^{1-1} - t_{20,1-1}^{1+1}
 + t_{20,1+1}^{1-1} - t_{20,1+1}^{1+1} )
 + \frac {i\sqrt{3}} { 2 }
                              (  t_{2-2,1-1}^{1-1} - t_{2-2,1-1}^{1+1}
                                + t_{2-2,1+1}^{1-1} - t_{2-2,1+1}^{1+1} 
	                        + t_{2+2,1-1}^{1-1} - t_{2+2,1-1}^{1+1}
                                + t_{2+2,1+1}^{1-1} - t_{2+2,1+1}^{1+1} ) ]$

45. $K_{xy,y}^{x'} =  \frac {3} {2} \times \frac {1} { 2 \sqrt{3} }
                                ( - t_{2-2,1-1}^{1-1} + t_{2-2,1-1}^{1+1}
                                   - t_{2-2,1+1}^{1-1} + t_{2-2,1+1}^{1+1} 
	                           + t_{2+2,1-1}^{1-1} - t_{2+2,1-1}^{1+1}
                                   + t_{2+2,1+1}^{1-1} - t_{2+2,1+1}^{1+1} )$

46. $\underline{ K_{xz,y}^{x'} } = - \frac {3} {2} \times \frac {i} { 2 \sqrt{3} }
                                   ( t_{2-1,1-1}^{1-1} - t_{2-1,1-1}^{1+1}
                              + t_{2-1,1+1}^{1-1} - t_{2-1,1+1}^{1+1} 
	                      - t_{2+1,1-1}^{1-1} + t_{2+1,1-1}^{1+1}
                              - t_{2+1,1+1}^{1-1} + t_{2+1,1+1}^{1+1} )$

47. $\underline{ K_{yz,y}^{x'} }=  \frac {3} {2} \times \frac {1} { 2 \sqrt{3} }
                              ( - t_{2-1,1-1}^{1-1} + t_{2-1,1-1}^{1+1}
                                   - t_{2-1,1+1}^{1-1} + t_{2-1,1+1}^{1+1} 
	                           - t_{2+1,1-1}^{1-1} + t_{2+1,1-1}^{1+1}
                                   - t_{2+1,1+1}^{1-1} + t_{2+1,1+1}^{1+1} )$

48. $\underline{ K_{zz,y}^{x'} } = - \frac {i} {  \sqrt{2} }
                                   ( t_{20,1-1}^{1-1} - t_{20,1-1}^{1+1}
                                   + t_{20,1+1}^{1-1} - t_{20,1+1}^{1+1} )$ 


49. $K_{xx,z}^{x'} =  \frac {1} {2} [
  -   ( t_{20,10}^{1-1} - t_{20,10}^{1+1} )
   +  \frac {\sqrt{3}} {2}  ( t_{2-2,10}^{1-1} - t_{2-2,10}^{1+1}
                              + t_{2+2,10}^{1-1} - t_{2+2,10}^{1+1} ) ]$

50. $K_{yy,z}^{x'} =  \frac {1} {2} [
  -  ( t_{20,10}^{1-1} - t_{20,10}^{1+1} ) 
 -\frac  {\sqrt{3}} {2} (  t_{2-2,10}^{1-1} - t_{2-2,10}^{1+1}
                               + t_{2+2,10}^{1-1} - t_{2+2,10}^{1+1} ) ]$

51. $\underline{ K_{xy,z}^{x'} }= - \frac {3} {2} \times \frac {i} { \sqrt{6} }
                                ( t_{2-2,10}^{1-1} - t_{2-2,10}^{1+1}
                                   - t_{2+2,10}^{1-1} + t_{2+2,10}^{1+1} )$

52. $K_{xz,z}^{x'} = \frac {3} {2} \times \frac {1}
                                   { \sqrt{6} } ( t_{2-1,10}^{1-1} - t_{2-1,10}^{1+1}
                              - t_{2+1,10}^{1-1} + t_{2+1,10}^{1+1} )$

53. $\underline{ K_{yz,z}^{x'} }= - \frac {3} {2} \times \frac {i} { \sqrt{6} }
                              ( t_{2-1,10}^{1-1} - t_{2-1,10}^{1+1}
                               + t_{2+1,10}^{1-1} - t_{2+1,10}^{1+1} )$

54. $K_{zz,z}^{x'} =    ( t_{20,10}^{1-1} - t_{20,10}^{1+1} )$ 


55. $K_{x,x}^{y'} = \frac {2} {3} \times \frac {i\sqrt{3} } { 4 }
                               ( t_{1-1,1-1}^{1-1} + t_{1-1,1-1}^{1+1}
                               - t_{1-1,1+1}^{1-1} - t_{1-1,1+1}^{1+1}
                               - t_{1+1,1-1}^{1-1} - t_{1+1,1-1}^{1+1}
                               + t_{1+1,1+1}^{1-1} + t_{1+1,1+1}^{1+1} )$

56. $\underline{ K_{y,x}^{y'} }= \frac {2} {3} \times \frac {\sqrt{3}} { 4  }
                               ( t_{1-1,1-1}^{1-1} + t_{1-1,1-1}^{1+1}
                              - t_{1-1,1+1}^{1-1} - t_{1-1,1+1}^{1+1}
                              + t_{1+1,1-1}^{1-1} + t_{1+1,1-1}^{1+1}
                              - t_{1+1,1+1}^{1-1} - t_{1+1,1+1}^{1+1} )$

57. $K_{z,x}^{y'} = \frac {2} {3} \times \frac {i} {2} \sqrt{\frac {3} {2} }
                              ( t_{10,1-1}^{1-1} + t_{10,1-1}^{1+1}
                     - t_{10,1+1}^{1-1} - t_{10,1+1}^{1+1} )$
		     
58. $\underline{ K_{x,y}^{y'} }= \frac {2} {3} \times \frac {\sqrt{3}} { 4 }
                     ( t_{1-1,1-1}^{1-1} + t_{1-1,1-1}^{1+1}
                             + t_{1-1,1+1}^{1-1} + t_{1-1,1+1}^{1+1}
                             - t_{1+1,1-1}^{1-1} - t_{1+1,1-1}^{1+1}
                             - t_{1+1,1+1}^{1-1} - t_{1+1,1+1}^{1+1} )$

59. $K_{y,y}^{y'} = - \frac {2} {3} \times \frac {i\sqrt{3}} { 4 }
                             (  t_{1-1,1-1}^{1-1} + t_{1-1,1-1}^{1+1}
                                 + t_{1-1,1+1}^{1-1} + t_{1-1,1+1}^{1+1}
                                 + t_{1+1,1-1}^{1-1} + t_{1+1,1-1}^{1+1}
                                 + t_{1+1,1+1}^{1-1} + t_{1+1,1+1}^{1+1} )$

60. $\underline{ K_{z,y}^{y'} }= \frac {2} {3} \times  
                                 \frac {1} {2} \sqrt{\frac {3} {2} }
                                 ( t_{10,1-1}^{1-1} + t_{10,1-1}^{1+1}
                    + t_{10,1+1}^{1-1} + t_{10,1+1}^{1+1} )$

61. $K_{x,z}^{y'} = \frac {2} {3} \times  \frac {i} {2} \sqrt{\frac {3} {2} }
                    ( t_{1-1,10}^{1-1} + t_{1-1,10}^{1+1}
                      - t_{1+1,10}^{1-1} - t_{1+1,10}^{1+1} )$

62. $\underline{ K_{y,z}^{y'} }=  \frac {2} {3} \times
                      \frac {1} {2} \sqrt{\frac {3} {2} }
                      ( t_{1-1,10}^{1-1} + t_{1-1,10}^{1+1}
                    + t_{1+1,10}^{1-1} + t_{1+1,10}^{1+1} )$

63. $K_{z,z}^{y'} = \frac {2} {3} \times \frac {i \sqrt{3} } { 2 }
                    ( t_{10,10}^{1-1} + t_{10,10}^{1+1} )$


64. $\underline{ K_{xx,x}^{y'} }= \frac {1} {2} [ 
 -  \frac {i} { \sqrt{2} }
                              ( t_{20,1-1}^{1-1} + t_{20,1-1}^{1+1}
                                - t_{20,1+1}^{1-1} - t_{20,1+1}^{1+1} )
 + \frac {i \sqrt{3}} { 2 }
                                ( t_{2-2,1-1}^{1-1} + t_{2-2,1-1}^{1+1}
                                - t_{2-2,1+1}^{1-1} - t_{2-2,1+1}^{1+1} 
	                        + t_{2+2,1-1}^{1-1} + t_{2+2,1-1}^{1+1}
                                - t_{2+2,1+1}^{1-1} - t_{2+2,1+1}^{1+1} ) ]$

65. $\underline{ K_{yy,x}^{y'} }=  \frac {1} {2} [ 
 -  \frac {i} { \sqrt{2} }
                              ( t_{20,1-1}^{1-1} + t_{20,1-1}^{1+1}
                                - t_{20,1+1}^{1-1} - t_{20,1+1}^{1+1} )
 - \frac {i \sqrt{3}} { 2 }
                                ( t_{2-2,1-1}^{1-1} + t_{2-2,1-1}^{1+1}
                                 - t_{2-2,1+1}^{1-1} - t_{2-2,1+1}^{1+1} 
	                         + t_{2+2,1-1}^{1-1} + t_{2+2,1-1}^{1+1}
                                 - t_{2+2,1+1}^{1-1} - t_{2+2,1+1}^{1+1} ) ]$

66. $K_{xy,x}^{y'} = \frac {3} {2} \times \frac {1} { 2 \sqrt{3} }
( t_{2-2,1-1}^{1-1} + t_{2-2,1-1}^{1+1}
                               - t_{2-2,1+1}^{1-1} - t_{2-2,1+1}^{1+1} 
	                       - t_{2+2,1-1}^{1-1} - t_{2+2,1-1}^{1+1}
                               + t_{2+2,1+1}^{1-1} + t_{2+2,1+1}^{1+1} )$

67. $\underline{ K_{xz,x}^{y'} } = \frac {3} {2} \times \frac {i} { 2 \sqrt{3} }
                               ( t_{2-1,1-1}^{1-1} + t_{2-1,1-1}^{1+1}
                                - t_{2-1,1+1}^{1-1} - t_{2-1,1+1}^{1+1} 
	                        - t_{2+1,1-1}^{1-1} - t_{2+1,1-1}^{1+1}
                                + t_{2+1,1+1}^{1-1} + t_{2+1,1+1}^{1+1} )$

68. $K_{yz,x}^{y'} =  \frac {3} {2} \times \frac {1} { 2 \sqrt{3} }
                                ( t_{2-1,1-1}^{1-1} + t_{2-1,1-1}^{1+1}
                              - t_{2-1,1+1}^{1-1} - t_{2-1,1+1}^{1+1} 
	                      + t_{2+1,1-1}^{1-1} + t_{2+1,1-1}^{1+1}
                              - t_{2+1,1+1}^{1-1} - t_{2+1,1+1}^{1+1} )$

69. $\underline{ K_{zz,x}^{y'} } =  \frac {i} {  \sqrt{2} }
                              ( t_{20,1-1}^{1-1} + t_{20,1-1}^{1+1}
                                - t_{20,1+1}^{1-1} - t_{20,1+1}^{1+1} ) $


70. $K_{xx,y}^{y'} =  \frac {1} {2} [ 
 - \frac {1} {  \sqrt{2} }
                              ( t_{20,1-1}^{1-1} + t_{20,1-1}^{1+1}
                                + t_{20,1+1}^{1-1} + t_{20,1+1}^{1+1} )
           +  \frac {\sqrt{3}} { 2 }
                                ( t_{2-2,1-1}^{1-1} + t_{2-2,1-1}^{1+1}
                                + t_{2-2,1+1}^{1-1} + t_{2-2,1+1}^{1+1} 
	                        + t_{2+2,1-1}^{1-1} + t_{2+2,1-1}^{1+1}
                                + t_{2+2,1+1}^{1-1} + t_{2+2,1+1}^{1+1} ) ]$

71. $K_{yy,y}^{y'} =  \frac {1} {2} [ 
  - \frac {1} {  \sqrt{2} }
                              ( t_{20,1-1}^{1-1} + t_{20,1-1}^{1+1}
                                + t_{20,1+1}^{1-1} + t_{20,1+1}^{1+1} )
- \frac {\sqrt{3}} { 2 }
                                ( t_{2-2,1-1}^{1-1} + t_{2-2,1-1}^{1+1}
                                 + t_{2-2,1+1}^{1-1} + t_{2-2,1+1}^{1+1} 
	                         + t_{2+2,1-1}^{1-1} + t_{2+2,1-1}^{1+1}
                                 + t_{2+2,1+1}^{1-1} + t_{2+2,1+1}^{1+1} ) ] $

72. $\underline{ K_{xy,y}^{y'} } = - \frac {3} {2} \times \frac {i} { 2 \sqrt{3} }
                                 ( t_{2-2,1-1}^{1-1} + t_{2-2,1-1}^{1+1}
                               + t_{2-2,1+1}^{1-1} + t_{2-2,1+1}^{1+1} 
	                       - t_{2+2,1-1}^{1-1} - t_{2+2,1-1}^{1+1}
                               - t_{2+2,1+1}^{1-1} - t_{2+2,1+1}^{1+1} )$

73. $K_{xz,y}^{y'}  = \frac {3} {2} \times \frac {1} { 2 \sqrt{3} }
                               ( t_{2-1,1-1}^{1-1} + t_{2-1,1-1}^{1+1}
                                + t_{2-1,1+1}^{1-1} + t_{2-1,1+1}^{1+1} 
	                        - t_{2+1,1-1}^{1-1} - t_{2+1,1-1}^{1+1}
                                - t_{2+1,1+1}^{1-1} - t_{2+1,1+1}^{1+1} )$

74. $\underline{ K_{yz,y}^{y'} } = - \frac {3} {2} \times \frac {i} { 2 \sqrt{3} }
                                ( t_{2-1,1-1}^{1-1} + t_{2-1,1-1}^{1+1}
                              + t_{2-1,1+1}^{1-1} + t_{2-1,1+1}^{1+1} 
	                      + t_{2+1,1-1}^{1-1} + t_{2+1,1-1}^{1+1}
                              + t_{2+1,1+1}^{1-1} + t_{2+1,1+1}^{1+1} )$

75. $K_{zz,y}^{y'} =   \frac {1} {  \sqrt{2} }
                              ( t_{20,1-1}^{1-1} + t_{20,1-1}^{1+1}
                                + t_{20,1+1}^{1-1} + t_{20,1+1}^{1+1} ) $


76. $\underline{ K_{xx,z}^{y'} }=  \frac {1} {2} [
 - {i} 
                              ( t_{20,10}^{1-1} + t_{20,10}^{1+1} )
    +  i \sqrt{ \frac {3} {2} }
                                ( t_{2-2,10}^{1-1} + t_{2-2,10}^{1+1}
                                + t_{2+2,10}^{1-1} + t_{2+2,10}^{1+1} ) ]$

77. $\underline{ K_{yy,z}^{y'} }=  \frac {1} {2} [
 - {i}
                              ( t_{20,10}^{1-1} + t_{20,10}^{1+1} )
-  i \sqrt{ \frac {3} {2} }
                                ( t_{2-2,10}^{1-1} + t_{2-2,10}^{1+1}
                                 + t_{2+2,10}^{1-1} + t_{2+2,10}^{1+1} ) ]$

78. $K_{xy,z}^{y'} = \frac {3} {2} \times \frac {1} { \sqrt{6} }
( t_{2-2,10}^{1-1} + t_{2-2,10}^{1+1}
                               - t_{2+2,10}^{1-1} - t_{2+2,10}^{1+1} )$

79. $\underline{ K_{xz,z}^{y'} } = \frac {3} {2} \times \frac {i} { \sqrt{6} }
                               ( t_{2-1,10}^{1-1} + t_{2-1,10}^{1+1}
                                - t_{2+1,10}^{1-1} - t_{2+1,10}^{1+1} )$

80. $K_{yz,z}^{y'} = \frac {3} {2} \times \frac {1} { \sqrt{6} }
                                ( t_{2-1,10}^{1-1} + t_{2-1,10}^{1+1}
                              + t_{2+1,10}^{1-1} + t_{2+1,10}^{1+1} )$

81. $\underline{ K_{zz,z}^{y'} } =   {i} 
                              ( t_{20,10}^{1-1} + t_{20,10}^{1+1} ) $


\clearpage

\begin{figure}
\includegraphics[scale=0.59]{fig1.eps} 
\caption{(color online) Polarization transfer coefficients
  from the doubly spin-polarized initial state nd 
  to the neutron  $K_{y,y}^{y'}$ and $K_{x,y}^{x'}$  
  in  elastic d(n,n)d scattering
  at $E=10, 70$, and $135$~MeV,
  shown as a function of c.m. angle $\Theta_{c,m.}$.
  The different lines are:
  the polarization transfer coefficient
  from the doubly spin-polarized initial state 
  to the neutron  ($K_{d,n}^{n'}$: black long dashed
  and green long-dashed-dotted lines), 
  its  constituent single polarization transfer coefficient 
  from the neutron to the neutron ($K_{0,n}^{n'}$: red solid
  and violet dotted lines) and 
  from the deuteron to the neutron  
  ($K_{d,0}^{n'}$: blue short-dashed and orange short-dashed-dotted lines).
  The red solid, blue short-dashed, and black long-dashed lines
  are predictions of the
  SMS chiral N$^4$LO$^+$ NN potential with regularisation parameter
  $\Lambda=450$~MeV. The violet dotted, orange short-dashed-dotted,
  and green long-dashed-dotted lines show effects of
  combining this potential with the 
  N$^2$LO chiral 3NF with the strength parameters $c_D=2.0$ and $c_E=0.2866$.
}
 \label{fig1}
\end{figure}

\begin{figure}
\includegraphics[scale=0.7]{fig2.eps}  
\caption{(color online) The same as in Fig.~\ref{fig1} but for  
  $K_{z,y}^{x'}$ and $K_{z,x}^{y'}$. The indigo solid and black dotted lines in
  d)-f) are the outgoing neutron induced polarizations $P_{y'}^{(0)}$
    predicted with NN and NN+3NF forces, respectively. 
}
 \label{fig2}
\end{figure}

\begin{figure}
\includegraphics[scale=0.7]{fig3.eps}  
\caption{(color online) The same as in Fig.~\ref{fig1} but for the 
  tensor polarized deuteron in the initial double polarized state 
  $K_{yy,y}^{y'}$ and $K_{xz,y}^{y'}$.
}
 \label{fig3}
\end{figure}

\begin{figure}
\includegraphics[scale=0.7]{fig4.eps}  
\caption{(color online)  Polarization transfer coefficients
  from the doubly spin-polarized initial state nd 
  to the neutron  $K_{y,y}^{y'}$ and $K_{x,y}^{x'}$
  in  kinematically complete breakup reaction d(n,n$_1$n$_2$)p  
  at $E=10, 70$, and $135$~MeV for exact FSI(1-3) condition, shown
  as a function of the 
  lab. angle $\theta_1^{lab}$ of the nucleon 1.  
  For description of lines see Fig.~\ref{fig1}.
}
 \label{fig4}
\end{figure}

\begin{figure}
\includegraphics[scale=0.7]{fig5.eps}  
\caption{(color online) The same as in Fig.~\ref{fig4} but for  
  $K_{z,y}^{x'}$ and $K_{z,x}^{y'}$.
}
 \label{fig5}
\end{figure}

\begin{figure}
\includegraphics[scale=0.68]{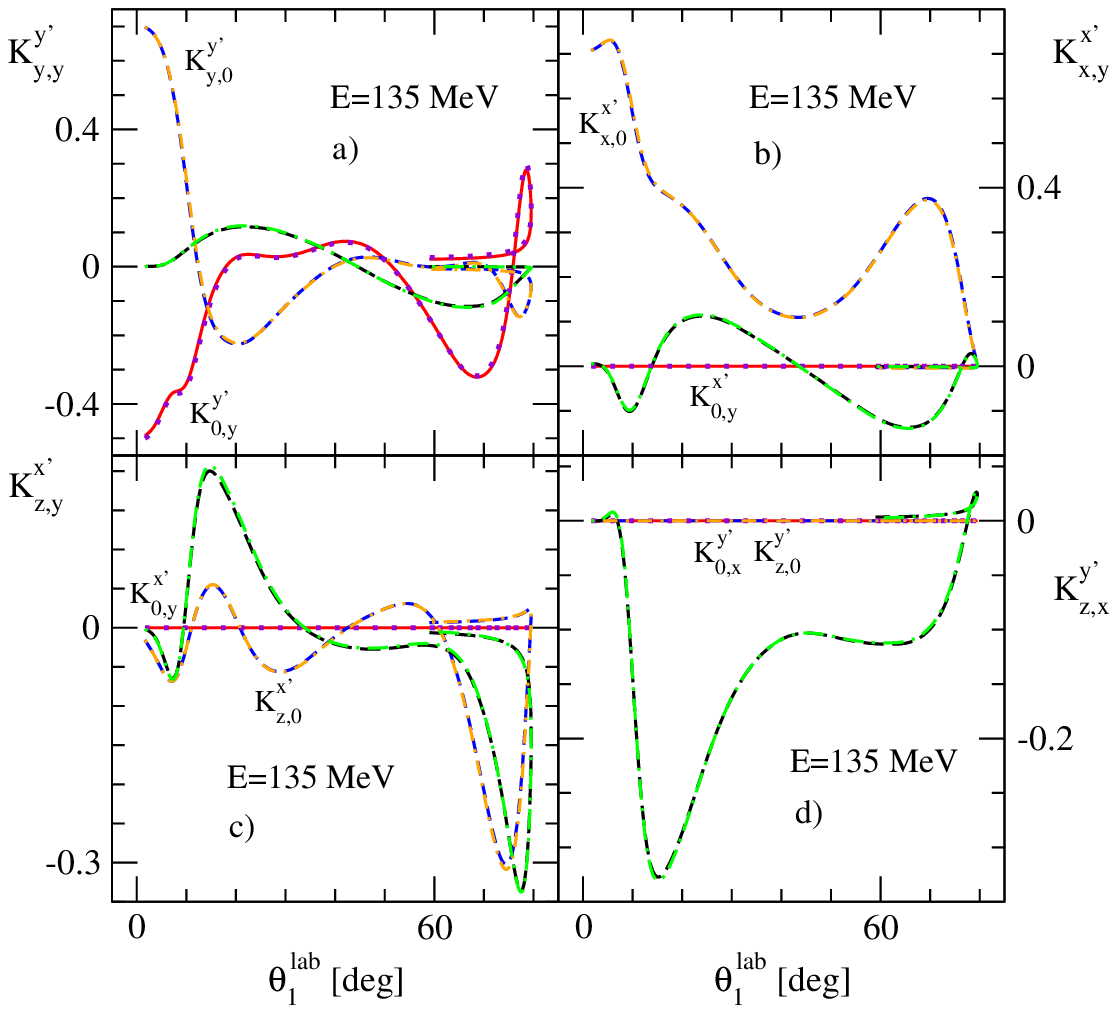}  
\caption{(color online)
  Polarization transfer coefficients
  in  kinematically complete breakup reaction d(n,n$_1$n$_2$)p  
  at $E=135$~MeV for exact QFS(1-2) condition, shown as a function of the 
  lab. angle  $\theta_1^{lab}$ of the nucleon 1.
  For description of lines see Fig.~\ref{fig1}.
}
 \label{fig6}
\end{figure}

\begin{figure}
\includegraphics[scale=0.7]{fig7.eps}  
\caption{(color online) 
  The polarization of the outgoing neutron $P_{y'}^n$ in $\vec d(\vec n,n)d)$
  elastic scattering at $E=135$~MeV for different signs combinations 
 of the  incoming neutron  and deuteron  polarizations 
 (orange short-dashed-dotted line): a) $p_y^n=+0.2$, $p_y^d=+0.6$, 
 b) $p_y^n=-0.2$, $p_y^d=-0.6$, c) $p_y^n=-0.2$, $p_y^d=+0.6$, and
 d) $p_y^n=+0.2$, $p_y^d=-0.6$.  
 Contributions  to $P_{y'}^n$ 
  from different terms in Eq.~(\ref{eqqq_1}) are given by the following lines:
  the term with induced polarization $P_{y'}^{(0)}$ - red solid,
  with single neutron transfer $K_{0,y}^{y'}$ - blue short-dashed,
  with single deuteron transfer $K_{y,0}^{y'}$ - black long-dashed, and
  with double-polarization transfer $K_{y,y}^{y'}$ - violet dotted.
}
 \label{fig7}
\end{figure}


\clearpage
                                

\begin{thebibliography}{99}

\bibitem{ohlsen1972} G. G. Ohlsen, Rep. Prog. Phys. {\bf{35}}, 717 (1972).

\bibitem{simonius1973} M. Simonius, Lecture Notes in Physics
  {\bf{30}}, 38 (1973).

\bibitem{przevoski} B. von Przevoski et al.,
  Phys. Phys. C {\bf{ 74}}, 064003 (2006).

\bibitem{kimsek}  K. Sekiguchi,  PoS(CD2021) 089 (2024).

\bibitem{Madison1971}
H. H. Barschall and W. Haeberli (Eds.),
\textit{Polarization Phenomena in Nuclear Reactions: Proceedings} (USA: Madison, University of Wisconsin Press, 1971).

\bibitem{glo96}  W. Gl\"ockle, H. Wita{\l}a, D. H\"uber, H. Kamada, J.
 Golak, Phys. Rep. {\bf{274}}, 107 (1996).
  
\bibitem{wit88}  H. Wita{\l}a, T. Cornelius and W. Gl\"ockle, 
  Few-Body Syst. {\bf{3}}, 123 (1988).
  
\bibitem{hub97} D. H\"uber, H. Kamada, H. Wita{\l}a, and W. Gl\"ockle, 
  Acta Physica Polonica {\bf B28}, 1677 (1997).
  
\bibitem{book} W. Gl\"ockle, The Quantum Mechanical Few-Body Problem,
  Springer Verlag 1983.

\bibitem{delt2005} A. Deltuva, A. C. Fonseca, and P. U. Sauer,
  Phys. Rev. C{\bf 72}, 054004 (2005).

\bibitem{wita2024_1} H. Wita{\l}a, J. Golak, R. Skibi\'nski,
  Phys. Rev. C {\bf 110}, 024005 (2024).

\bibitem{wita2024_2} H. Wita{\l}a, J. Golak, R. Skibi\'nski,
  Phys. Rev. C {\bf 110}, 014003 (2024).

\bibitem{brinksatch} D. M. Brink and G. R. Satchler, Angular Momentum,
 Clarendon Press Oxford 1993.
  
\bibitem{reinert2018} P. Reinert, H. Krebs, and E. Epelbaum,
  Eur. Phys. Journal A{\bf{54}}, 86  (2018).
  
\bibitem{maris2021} P. Maris et al. (LENPIC Collaboration),
  Phys. Rev. C {\bf 103}, 054001 (2021).

\bibitem{vankolck} U. van Kolck, Phys. Rev. C {\bf 49}, 2932 (1994).

\bibitem{epel2002} E. Epelbaum, A. Nogga, W. Gl\"ockle, H. Kamada,
  Ulf-G. Mei{\ss}ner, and H. Wita{\l}a, Phys. Rev. C {\bf 66}, 064001 (2002).

\bibitem{epel2020} E. Epelbaum et al., Eur. Phys. Journal A{\bf{56}}, 92 (2020).

\bibitem{wita1998} H. Wita{\l}a, W. Gl\"ockle, D. H\"uber, J. Golak,
  and H. Kamada, Phys. Rev. Lett. 81, 1183 (1998).
  
\bibitem{wita2001} H. Wita{\l}a, W. Gl\"ockle, J. Golak, A. Nogga,
  H. Kamada, R. Skibi\'nski and J. Kuro\'s-\.Zo{\l}nierczuk,
  Phys. Rev. C {\bf 63}, 024007 (2001), and references therein.

\bibitem{shim1982} H. Shimizu et al., Nucl. Phys. A{\bf{382}}, 242 (1982).

\bibitem{sekiguchi_dtarget} K. Sekiguchi et al., 
  Phys. Rev. C {\bf 96}, 64001 (2017).

\bibitem{tateishi_ptarget1} K. Tateishi, Proc. Nat. Acad. Sci.
  {\bf 111}, 7527 (2014).
  
\bibitem{sakaguchi_ptarget2} S. Sakaguchi et al., 
  Phys. Rev. C {\bf 87}, 021601(R) (2013).

\end{thebibliography}
\end{document}